\documentclass[usenatbib,a4paper]{mn2e}
\usepackage{graphicx,natbib,amssymb,amsmath,times,url}

\newcommand{\gsim}{\mathrel{\hbox{\rlap{\lower.55ex \hbox {$\sim$}}
                   \kern-.3em \raise.4ex \hbox{$>$}}}}
\newcommand{\lsim}{\mathrel{\hbox{\rlap{\lower.55ex \hbox {$\sim$}}
                   \kern-.3em \raise.4ex \hbox{$<$}}}}

\title[Inefficient star formation]{Inefficient star formation: The combined effects of magnetic fields and radiative feedback}

\author[Price \& Bate]{Daniel J. Price$^{1,2}$ and Matthew R. Bate$^{2}$ \\
$^{1}$Centre for Stellar and Planetary Astrophysics, School of Mathematical Sciences, Monash University, Clayton Vic 3168, Australia\\
$^{2}$School of Physics, University of Exeter, Stocker Rd, Exeter EX4 4QL, UK \\
}
\date{Submitted: 15th Jan 2009 Revised: 6th April 2009 Accepted: 24th April 2009}
\pagerange{\pageref{firstpage}--\pageref{lastpage}} \pubyear{2009}

\begin{document}
\label{firstpage}
\bibliographystyle{mn2e}
\maketitle

\begin{abstract}
 We investigate the effects of magnetic fields and radiative protostellar feedback on the star formation process using self-gravitating radiation magnetohydrodynamical calculations.  We present results from a series of calculations of the collapse of $50~{\rm M}_{\odot}$ molecular clouds with various magnetic field strengths and with and without radiative transfer. 
 
 We find that both magnetic fields and radiation have a dramatic impact on star formation, though the two effects are in many ways complementary. Magnetic fields primarily provide support on large scales to low density gas, whereas radiation is found to strongly suppress small-scale fragmentation by increasing the temperature in the high-density material near the protostars. With strong magnetic fields and radiative feedback the net result is an inefficient star formation process with a star formation rate of $\lesssim 10$\% per free-fall time that approaches the observed rate, although we have only been able to follow the calculations for ~1/3 of a free-fall time beyond the onset of star formation.
\end{abstract}


\section{Introduction}
 Star formation is a remarkably inefficient process. This inefficiency in itself is a very good thing for the universe as a whole, since without it galaxies such as the Milky Way would very quickly exhaust their supplies of gas by converting it into stars. Recent estimates from the c2d \emph{Spitzer} legacy survey of five nearby molecular clouds suggest that around $3-6$\% of the available gas in a molecular cloud is converted into stars in the local region of our Galaxy \citep{evansetal09}. Previous observational results suggest similarly low efficiencies: e.g., $\approx 1-6$\% in Taurus \citep{el91,onishietal98};  $<13$\% in the clouds in Chamaeleon \citep{mizunoetal99}, though some dispersion in these results arises from the use of differing measures, whereas \citeauthor{evansetal09} use a uniform definition of efficiency for all clouds.
 
  The source of such uniformly low efficiency is poorly understood, and it remains unclear as to what the ``rate-limiting step'' in star formation really is, since inefficiency is apparently present at all levels, from the formation of molecular clouds in galaxies \citep{dobbsetal08,leroyetal08} to the fact that only small, clustered regions of molecular clouds with mass fractions of $\lsim 20$ percent (\citealt{lada92}; \citealt*{jdk04}; \citealt{hatchelletal05}) participate in star formation, to the observation that only a fraction of the mass in dense molecular cloud cores ends up as stars \citep{bm89,alvesetal07}.    Inspite of this, it is clear that a large part of the inefficiency lies within molecular clouds themselves.
  
 From a theoretical perspective, we have a very good idea of the basic ingredients of the star formation process - namely gravity, gas dynamics, turbulence, magnetic fields, radiative and mechanical feedback, though their relative importance (particularly with respect to magnetic fields and turbulence) remains vigourously debated \citep[e.g.][]{cht09,mt08,cht08b}. By definition star formation involves the conversion of gas into stars under self-gravity, the basics of which were elucidated by \citet{jeans02}. The complication to the gas dynamics is the highly turbulent (and supersonic) nature of molecular clouds and the wide range of length and time scales over which star formation takes place, presenting a formidable challenge for numerical simulations even before considering other relevant physics. Nevertheless, simulations including just self-gravity and hydrodynamics \citep*{kbb98,bbb02a,bbb02b,bbb03,bbv03,bb05,bate05} have been surprisingly successful in predicting many properties of clustered star formation, including the initial mass function (though with an overproduction of brown dwarfs), multiplicity as a function of primary mass, the frequency of very low mass binaries, general trends for the separation and mass ratio distributions of binaries and the relative orbital orientations of triple systems \citep{bate09a}. 
 
However, the efficiency of star formation in these calculations would be $\gsim 50$\% were the simulations left to run, since in the absence of stellar feedback there can be nothing to prevent all of the (bound) gas from eventually accreting onto the stars. While tidal forces from the underlying galactic potential may be a contributing factor in some cases \citep{bpetal09}, many suggestions in the literature are related to the turbulence present in the cloud.  Star formation in initially unbound clouds, where star formation will nevertheless occur in the presence of a turbulent velocity field \citep{cb04,clarketal05}, is one extreme.  But sources of turbulent driving  \citep*[e.g.][]{mm00,kmm06,nl07}, whilst not strictly changing the overall efficiency in a bound cloud, can dramatically alter the fraction of a cloud which is unstable to gravitational collapse in a given dynamical time, decreasing the efficiency per free-fall time  \citep[e.g.][]{padoan95,khm00,km05}. This occurs naturally in turbulent models because there is a spectrum of density fluctuations, of which only a small fraction is in sub-regions dense enough to be Jeans-unstable. Star formation is thus made inefficient, in a per free-fall time sense, because turbulence produces in a given dynamical time a range of clumps (or ``cores''), only some of which are bound and will collapse \citep{khm00}. There is still a difficulty, however, which is that if this is to work for dense, globally bound regions, one has to keep driving the turbulent motions (either from outside or within) otherwise the turbulence will quickly decay and the gas from the whole dense globally bound clump will eventually be used up in star formation.


  Magnetic fields have long been recognised as a key ingredient in star formation \citep{ms56,sal87,mestel99}, given that observations robustly measure fields at sufficient strengths that they are close to preventing star formation altogether \citep{mk04}, and robustly in the regime where magnetic pressure is dominant over gas pressure \citep{crutcher99,bourkeetal01,ht05}. The importance of the latter point is easily overlooked and implies that, even if magnetic fields do not prevent global gravitational collapse in a molecular cloud, they can nevertheless act as the dominant source of pressure \citep{pb08}, supporting large fractions of the cloud and perhaps regulating star formation \citep{nl05}. Magnetic fields, may also have important effects on the statistics of molecular cloud turbulence, even in the regime where the Alfv\'en speed is small compared to the turbulent velocities \citep{padoanetal07}. Importantly, magnetic fields are not usually included in determinations of whether or not a molecular cloud core is ``gravitationally bound''.
  
   Radiation presents a complementary method for regulating star formation and the need for star formation simulations to incorporate the effects of radiative transfer has also long been understood \citep[e.g.][]{larson69,bb75,bm92,mi00}. Radiative feedback affects star formation as soon as the gas becomes optically thick, setting the ``opacity limit'' beyond which fragmentation can proceed no further \citep{ll76,rees76}. From thereon the newborn protostar can continue to radiate into the surrounding gas, increasing the Jeans mass and thus inhibiting further star formation \citep{wb06,krumholz06,bate09b}. In the case of massive stars, radiation may be sufficient to halt accretion from the cloud \citep{kahn74,wc87} although various non-spherical and time-dependent effects mitigate this effect \citep*{nakano89,nhn95, ja96,ys02,kmk05,kkm07}.

 However, neither magnetic fields nor radiation are easy to incorporate into three-dimensional numerical simulations of the star formation process. The development of algorithms for magnetohydrodynamics (MHD) \citep{pm04a,pm04b,pm05,pb07} and radiative transfer in the flux-limited diffusion approximation \citep*{wb04,wb05,wb06} in the context of the smoothed particle hydrodynamics (SPH) method has nevertheless paved the way for such effects to be included. Thus we have recently been able to incorporate, though separately, the effects of magnetic fields \citep{pb08} and radiative feedback \citep{bate09b} into simulations of star cluster formation. In this paper we study, for the first time, the combined effects of both.
 
  We thus present a `recipe' for inefficient star formation in even relatively dense molecular clouds. The ingredients (i.e., the equations of self-gravitating radiation MHD and our numerical formulation of them) are presented in Section \ref{sec:numerics}. The initial conditions for our simulations are discussed in Section \ref{sec:ics}. We present our results in Section \ref{sec:results} and discuss their wider implications in Section \ref{sec:discussion}.

\section{Numerical method}
\label{sec:numerics}

 In this paper we solve the equations of self-gravitating radiation magnetohydrodynamics, using a two-temperature flux-limited diffusion scheme for the radiation, coupled with the equations of ideal MHD (that is, assuming infinite conductivity and without considering ambipolar (ion-neutral) diffusion or the Hall effect). Whilst we have previously published star cluster formation calculations using separately either the MHD \citep{pb07,pb08} or using the flux-limited diffusion \citep{wb06,bate09b} schemes, this is the first time which we have combined the two. Thus, whilst the MHD formulation is identical to that used in \citet{pb07} and \citet{pb08} and the radiation scheme is based on that used in \citet{bate09b}, some minor changes have been made to the radiation terms in order to combine them with the MHD part of the code.

\subsection{Equations of Radiation Magnetohydrodynamics}
 The equations of self-gravitating radiation MHD are solved in the form
\begin{eqnarray}
\rho & = & \int \delta({\bf r} - {\bf r}') \rho' dV', \label{eq:cty} \\
\frac{d{\bf v}}{dt} & = & -\frac{1}{\rho}\nabla\left(P + \frac12 \frac{B^{2}}{\mu_{0}} - \frac{{\bf B}{\bf B}}{\mu_{0}}\right) + \frac{\chi}{c}{\bf F} - \nabla\Phi, \label{eq:mom} \\
\frac{du}{dt} & = & -\frac{P}{\rho}\nabla\cdot{\bf v} + ac\kappa \left[\frac{\rho\xi}{a} - \left(\frac{u}{c_{v}}\right)^{4} \right], \label{eq:dudt} \\
\frac{d\xi}{dt}  & = & -\frac{\nabla\cdot{\bf F}}{\rho} - \frac{\nabla{\bf v:P}_{rad}}{\rho}-ac\kappa \left[\frac{\rho\xi}{a} - \left(\frac{u}{c_{v}}\right)^{4} \right], \label{eq:dxidt} \\
{\bf B} & = & \nabla\alpha_{E} \times \nabla\beta_{E}, \label{eq:B} \\
\frac{d\alpha_{E}}{dt} & = & 0; \hspace{1cm} \frac{d\beta_{E}}{dt} = 0. \label{eq:euler} \\
\nabla^{2}\Phi & = & 4\pi G\rho, \label{eq:grav}
\end{eqnarray}
where $\rho$ is the density, ${\bf v}$ is the velocity, $P$ is the hydrodynamic pressure, ${\bf B}$ is the magnetic field, $u$ is the specific thermal energy of the gas, $\Phi$ is the gravitational potential, $\xi$ and ${\bf P}_{rad}$ are the frequency-integrated specific radiation energy and radiation pressure tensor respectively; $a$, $c$, $\chi$, $\kappa$ and $c_{v}$ are the radiation constant, the speed of light, the total and absorption opacities and the ratio of specific heats respectively and ${\bf F}$ is the radiative flux, which in the flux-limited diffusion approximation is given by:
\begin{equation}
{\bf F} = \frac{c\lambda}{\kappa\rho}\nabla(\rho \xi),
\end{equation}
where $\lambda$ is the dimensionless flux limiter which is designed to ensure that the radiation propagates no faster than the speed of light (see \citealt{wb05} for details). The above expression for ${\bf F}$ means that the first term in equation (\ref{eq:dxidt}) becomes a diffusion term for the radiation energy (hence ``flux-limited diffusion'').

Equation (\ref{eq:cty}) is an exact solution to the continuity equation which is represented in SPH form by the density summation (see \citealt{price08} for the difference between integral and differential formulations in an SPH context). Equation (\ref{eq:mom}) is the equation of motion for the gas which contains force terms from the hydrodynamic ($\nabla P$), magnetohydrodynamic $\nabla\left[\frac{1}{\mu_{0}} \left(\frac12 B^{2} - {\bf B}{\bf B}\right)\right]$ and radiation $({\bf F}$, ie. $\nabla(\rho\xi))$ pressure gradients and from the gradient in the gravitational potential ($\nabla\Phi$). Equations (\ref{eq:dudt}) and (\ref{eq:dxidt}) are the energy equations for the gas and radiation respectively. Equation (\ref{eq:B}) is an expression of the magnetic field in terms of the Euler (or Clebsch) potentials $\alpha_{E}$ and $\beta_{E}$ which maintains the divergence constraint ($\nabla\cdot{\bf B} = 0$) by construction and for which the induction equation for the magnetic field takes the particularly simple form given by equation ($\ref{eq:euler}$) \citep{stern70,rp07}.
  It should be noted that use of the Euler potentials approach also introduces limitations on the topology of fields that can evolve during the calculation. Whilst these are discussed in more detail in \citet{pb08} and \citet{pb07}, the main physical process not captured is the winding up of magnetic fields on smaller scales, since the Euler potentials rely on a well defined mapping from the initial particle positions to those at a later time. This means that, whilst we are able to study the influence of magnetic fields on the large scale structure of the cloud, field growth on smaller scales is not well captured. On the other hand the ideal MHD approximation also breaks down at these scales, so an improved formulation would also need to correctly account for non-ideal MHD effects such as resistivity and ambipolar diffusion.
Finally, Poisson's equation (Equation \ref{eq:grav}) is solved in order to determine the gravitational force.

 The equation set is closed by equations of state for the gas and the radiation field. For the gas, the equation of state is given by the ideal gas law
\begin{equation}
P = \frac{\rho \mathcal{R} T}{\mu}, \label{eq:gameos}
\end{equation}
where $T$ is the gas temperature, $\mathcal{R}$ is the gas constant and $\mu$ is the mean molecular weight. The equation of state takes into account the translational, rotational and vibrational degrees of freedom of molecular hydrogen (assuming a 3:1 mix of ortho- and para- hydrogen that remains fixed throughout the calculations; see \citealt{boleyetal07}). It also includes the dissociation of molecular hydrogen and the ionisations of hydrogen and helium (which are assumed to have mass fractions of $X=0.7$ and $Y=0.28$ respectively). The contributions of metals to the equation of state is neglected.

For the radiation, the equation of state is given by the Eddington approximation
\begin{equation}
{\bf P}_{rad} = {\bf f}\rho\xi,
\end{equation}
where ${\bf f}$ is the Eddington tensor which has both an isotropic term and an anisotropic term related to the gradient in radiation energy density (see \citealt{wb06} for details).

For comparison with previous results, we have also performed a set of calculations without radiative transfer, but which use a barotropic equation of state for the gas (i.e. replacing equations~\ref{eq:dudt}, \ref{eq:dxidt} and \ref{eq:gameos}) of the form
\begin{equation}
P = K \rho^{\gamma}.
\end{equation}
where the polytropic exponent $\gamma$ is given by
\begin{eqnarray}
\gamma = 1,  & & \rho \le 10^{-13} {\rm g\phantom{l}cm}^{-3}, \nonumber \\
\gamma = 7/5, & &  \rho > 10^{-13} {\rm g\phantom{l}cm}^{-3}.
\label{eq:eos}
\end{eqnarray}
 The simulations using the barotropic equation of state are thus identical to those performed by \citet{pb08} except for a factor-of-ten decrease in the sink particle radii (see below) and also the MPI-parallelisation of the tree-code, both of which change the overall fragmentation pattern slightly due to the chaotic nature of star formation.

\subsection{Numerical method}
 We solve equations (\ref{eq:cty})--(\ref{eq:grav}) using the Smoothed Particle Hydrodynamics (SPH) method (for reviews see \citealt{monaghan92,price04,monaghan05}). The SPH formulations of various parts of these equations, as currently implemented in our code, have been separately described and tested in a number of papers (mostly involving the authors), as summarised below.

\subsubsection{SPH formulation}
  The self-gravitating part of the algorithm (i.e. equation (\ref{eq:grav}) and the gravitational force term in equation (\ref{eq:mom})) is identical to the energy-conserving formulation described and tested in \citet{pm07}.  The gravitational force is softened using a softening length that is equal to the SPH smoothing length and formulated such that taking the Laplacian of the gravitational potential results in precisely the right hand side of Poisson's equation (\ref{eq:grav}) with the density $\rho$ equal to that calculated in the hydrodynamics via the SPH summation (ie. the SPH expression of equation~\ref{eq:cty}) \citep{pm07}. Furthermore -- despite the softening length being a variable function of position -- momentum, energy and angular momentum are conserved exactly using this formalism. However, in practice, a nearest-neighbour binary tree algorithm is used to efficiently calculate the long-range part of the gravitational force (and also return the list of SPH neighbours), which does not conserve momentum, angular momentum or energy exactly.  The tree code formed the original core of the SPH code and remains essentially as originally implemented by \citet{benzetal90}.
  
   All of the evolution equations are integrated using a second order Runge-Kutta-Fehlberg method with a timestep related to the convergence of each variable with timestep (that is, determined by comparing the error using half of the current timestep to that over the full timestep). Given the rich array of physics in our current calculations, we have found this to be substantially more accurate than a standard leapfrog method where the timestep is based only on stability considerations (rather than accuracy). Individual particle timesteps were implemented by \citet{bate95} in order to efficiently follow calculations where the timestep is constrained by only a small fraction of particles in a simulation.

 The hydrodynamics and MHD parts of the code (i.e., the numerical formulation of equations (\ref{eq:cty})-(\ref{eq:dudt}) apart from the radiation terms and equations (\ref{eq:B})-(\ref{eq:euler})) are based on the smoothed particle magnetohydrodynamics algorithm developed by \citet{pm04a,pm04b,pm05} and applied to star formation using the Euler potentials formulation by \citet{pb07,pb08} (see also \citealt{rp07}). Special attention has been paid to the formulation of terms relating to the gradient of smoothing length \citep{pm04b,pm07} which ensures that total energy and entropy are conserved exactly by the hydrodynamic parts of the equations. {Energy and momentum conservation is not maintained exactly for the magnetic parts of the equations in order to avoid the well-known instability relating to exactly momentum-conserving formulations of the SPMHD force term (see \citealt{pm05}). Furthermore, using the Euler potentials, the force equation is not directly derived from the numerical form of the induction equations leading to a very small error in energy conservation. In practice, however, these errors are much smaller than those introduced by the treecode for the gravitational force and individual particle timesteps.
 
 Dissipative terms corresponding to artificial viscosity and artificial resitivity are added in order to capture shocks and magnetic reconnection, respectively. These are applied as described in \citet{pm05} and for the Euler potentials by \citet{pb07} and \citet{rp07}. For the calculations with radiative transfer, the energy associated with this dissipation is added to the thermal energy, though obviously such energy is discarded for the calculations employing a barotropic equation of state and thus also during the initial period ($t < 1 t_{ff}$) for all the calculations during which the barotropic equation of state has been used (see below). We are, therefore, not able to realistically assess the effect of any heating that may arise due to magnetic reconnection since any heat created by reconnection in the early phases is lost and at later times, whilst the energy is captured, the field structure is effectively lost because of the limitations to the Euler potentials approach on smaller scales (see above). Thus, the contribution of magnetic dissipation to heating in the present calculations is very small.

 The radiative transfer parts of equations (\ref{eq:mom}) and (\ref{eq:dudt}) and the radiative energy equation (\ref{eq:dxidt}) are solved implicitly using the formulation developed by \citet{wb04}, accelerated by \citet{wb05} and applied to star formation by \citet{wb06} and \citet{bate09b}.  We use the same opacities as \cite{wb06}.  In order to combine the radiative transfer parts of the code (developed by \citealt{wb05} and based on a traditional ``number of neighbours'' approach to variable smoothing lengths in SPH) with the MHD (developed by \citealt{pm04b}, which is formulated taking account of variable smoothing length gradient terms), minor modifications have been made to the manner in which the radiative transfer equations are expressed in SPH form. The main change is that the radiation diffusion term in equation (\ref{eq:dxidt}) is calculated using an average of the kernels rather than an average smoothing length as in \citet{wb06}, so that the diffusion term in the energy equation becomes
\begin{equation}
\left(\frac{d\xi_{i}}{dt}\right)_{\rm diff} = \sum_{j} \frac{m_{j}c}{\rho_{i}\rho_{j}}\left[\frac{4 D_{i} D_{j}}{D_{i} + D_{j} } \right]\left(\rho_{i}\xi_{i} - \rho_{j}\xi_{j}\right) \frac{\overline{\nabla W_{ij}}}{r_{ij}}
\end{equation}
where the average of the SPH kernel gradients is
\begin{equation}
\overline{\nabla W_{ij}} =  \frac12\left[\nabla W_{ij}(h_{i}) + \nabla W_{ij}(h_{j})\right],
\end{equation}
and 
\begin{equation}
D_i = \frac{\lambda_{i}}{\kappa_{i}\rho_{i}}.
\end{equation}
 Note that in the variable smoothing length formulation of SPH, the smoothing length $h$ is an analytic function of the density $\rho$, which is in turn a function of smoothing length via the SPH density summation. A solution to the density summation must therefore be obtained iteratively as described in \cite{pm07}.



\subsection{Sink particles}
 Sink particles were introduced into SPH by \citet*{bbp95} in order to follow star formation calculations beyond the formation of the first star. 
 
For the calculations presented here that do not include radiative transfer, the criterion used for sink particle creation is identical to that described in \citet{pb08} and we therefore refer the reader to that paper for details. The major difference between the calculations of \citeauthor{pb08} and the similar calculations in this paper is that we have used sink particles with an accretion radius of only $0.5$ AU, compared to $5$ AU in \citet{pb08} (and similarly in BBB03). This adds considerable computational expense to the calculations because the closest gas orbits around the sinks (and the smallest length scale $\sim h$) are reduced by a factor of 10, resulting in an increase in the maximum acceleration by a factor of $1/r^{2} = 100$ and therefore a decrease in the minimum timestep in the calculations $\Delta t \propto \sqrt{h/\vert a \vert}$ by a factor of $\sim$30. Whilst such expense is unnecessary when using a barotropic equation of state (in Section \ref{sec:results} we compare our results to previous results obtained by \citealt{pb08} using 5~AU sink radii and find essentially no difference to the fragmentation), it is important for the radiative transfer calculations.

With the exception of the addition of magnetic fields, the radiative transfer calculations presented here are very similar to those recently published by \citet{bate09b}.  In both, the gas is followed beyond the first hydrostatic core phase \citep{larson69} and the onset of molecular hydrogen dissociation ($T\approx 2000$ K). Sink particles are inserted during the second collapse phase, just before a stellar core would be formed in the calculations.  \citet{bate09b} inserted sink particles at a density of $10^{-5}$ g~cm$^{-3}$ while in the calculations presented here they are inserted slightly earlier at $10^{-6}$ g~cm$^{-3}$.  In terms of the real star formation process, this is just a couple of weeks before the stellar core is formed.  As in \citet{bate09b}, no radiative feedback is provided by the sink particle.  The radiative feedback provided by the protostars is limited to the radiation emitted from the gas as it falls into the sink particles. Thus, it is important to make the sink particle accretion radii as small as is computationally practical (0.5 AU in both the calculations of \citealt{bate09b} and the calculations presented here).  As noted by \citet{bate09b}, because not all of the protostellar luminosity is fed back into the calculations, the effects of radiative feedback seen in the calculations presented here must be viewed as a lower limit.


An estimate of the energy input that we are missing from accretion within the sink particle radius can be made by comparing the accretion luminosity from within this region to that available from infinity. The accretion energy expected from within our sink particle radius of 0.5~AU is given by
\begin{equation}
L_{acc} = GM\dot{M} \left(\frac{1}{R_{*}} - \frac{1}{0.5 {\rm AU}} \right).
\end{equation}
This may be compared to the accretion luminosity that \emph{is} captured in our calculations by accretion to the sink radius from infinity, which is given by $L_{acc} = GM\dot{M}/(0.5~{\rm AU})$.  Therefore, if we assume a protostellar radius of $\sim 3 R_{sun}$ there is potentially a factor of up to $\sim 30$ in further energy input that is missing from the current calculations.  \cite{bate09b} investigated the effect of this missing radiation on his similar calculations that did not include magnetic fields by repeating a calculation with a larger accretion radius of 5~AU (i.e. reducing the accretion luminosity by a further order of magnitude).  He found only a small difference in the amount of fragmentation that occurred between the 0.5~AU and 5~AU calculations because even the heating present in the calculation with 5-AU accretion radii was enough to inhibit fragmentation near to existing protostars.  Thus, while we again emphasize that the radiative feedback incorporated into the current calculations is only a lower limit, we believe that using accretion radii of 0.5~AU captures the essence of the effects of radiative feedback, at least in terms of fragmentation.  This is in stark contrast to the situation encountered using a barotropic equation of state.

\begin{figure*}
   \centering
   \includegraphics[angle=270,width=\textwidth]{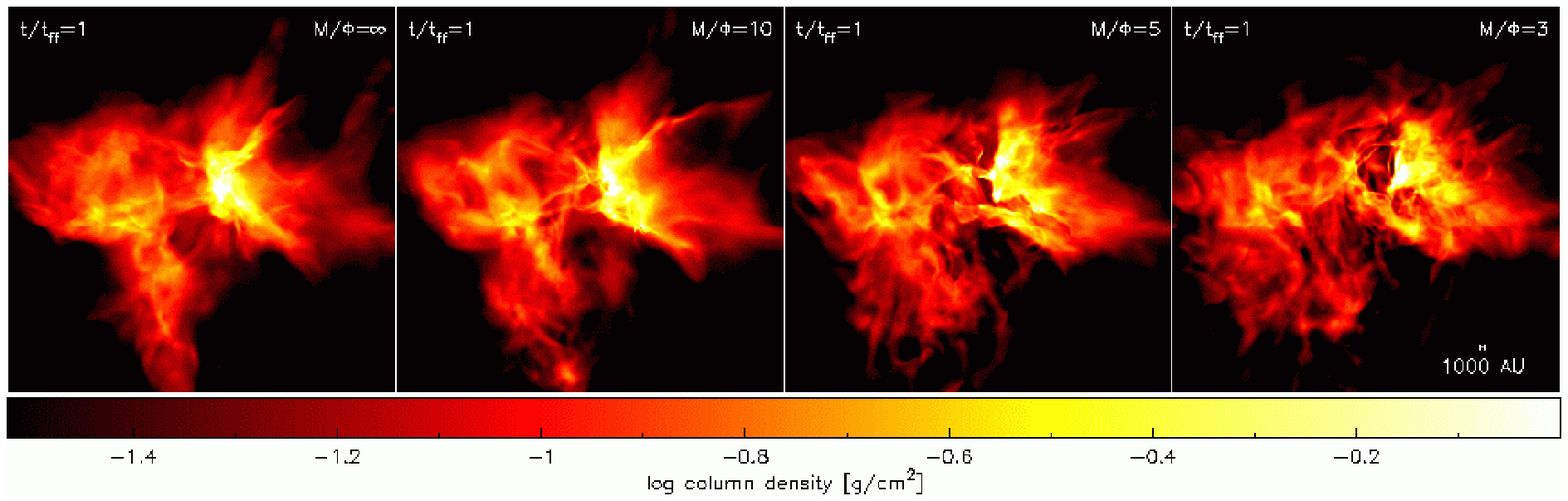}
   \caption{Global cloud structure at 1.0 initial cloud free-fall time, for progressively increasing magnetic field strength [mass-to-flux ratios of $\infty$ (that is, hydrodynamic), $10$, $5$ and $3$ in units of the critical value], showing the dramatic influence of magnetic fields on the large-scale structure of the cloud.  As in \citet{pb08}, the magnetic field has a dominant influence in the regime where $\beta < 1$ (third and fourth panels), producing large-scale magnetic-pressure-supported voids  and column density structures aligned with the magnetic field in the cloud envelope. Note that radiative feedback plays no role at this stage - the cloud structure is determined entirely by the interaction between turbulence, gravity and the large scale magnetic field.}
   \label{fig:global}
\end{figure*}

\section{Initial conditions}
\label{sec:ics}

 The initial conditions for the simulations are identical to those presented by \citet{pb08} and similar to the original calculation of \cite{bbb03} and the first of the calculations performed by \citet{bate09b}. We briefly recap the initial conditions below.

\subsection{Density, temperature and velocity field}
  We set up an initially uniform, spherical cloud with a diameter of $0.375$ pc (77,400 AU) that contains a total of 50 ${\rm M}_{\odot}$ of molecular gas, giving an initial density of $\rho_{0} = 1.2 \times 10^{-19}$g cm$^{-3}$ ($n_{\rm H_{2}} = 3\times 10^{4}$) and a global free-fall time of $t_{\rm ff} = \sqrt{3\pi/(32\rho_{0} G)} = 1.90\times 10^5$ yrs.

 The cloud is constructed using 3.5 million SPH particles (determined by the resolution requirement for resolving the Jeans mass by \citealt{bateburkert97}, see \citealt{bbb03}) placed in a uniform random distribution cropped to the cloud radius (i.e. no particles are placed exterior to the cloud).  This results in a significant expansion of the outer layers as the calculation proceeds (equivalent to the assumption of open boundary conditions in a grid-based simulation). The initial sound speed was set to $1.84 \times 10^{4}$ cm~s$^{-1}$, corresponding to a temperature of $10$~K given the mean molecular weight of $\mu = 4.0/(2\times 0.7+0.28) = 2.38$ amu. The resultant ratio of thermal to gravitational energy was $\alpha_{\rm grav} = 0.074$.
 
A supersonic `turbulent' velocity field with power spectrum $P(k)\propto k^{-4}$ (i.e. consistent with Larson's scaling relations, \citealt{larson81}) was imposed upon the initially uniform density cloud as in \citet{bbb03}, with the initial velocity field normalised such that the kinetic energy is initially equal to the gravitational potential energy of the cloud.  This gives an initial root mean square (RMS) Mach number of 6.4 and an initial RMS velocity of $1.17 \times 10^{5}$ cm/s.

 The computational challenge of star formation is well illustrated by the fact that during the calculations, we find that the densest regions can contain particles moving on a timestep up to $2^{19}$ times smaller than the largest timestep bin (which is constrained by the time between output dumps), so that the shortest timestep is around $1.5$ hours compared to a total evolution time of several hundred thousand years.

\subsection{Magnetic fields}
We quantify the relative strength of the magnetic field in terms of the mass-to-flux ratio ($M/\Phi$) of the cloud, compared to the critical value for the onset of collapse in a spherical cloud given by \citep[e.g.][]{ms76,mestel99,mk04}
\begin{equation}
\left(\frac{M}{\Phi}\right)_{\rm crit} = \frac{2 c_{1}}{3} \sqrt{\frac{5}{\pi G \mu_{0} }},
\label{eq:mphicrit}
\end{equation}
where $G$ and $\mu_{0}$ are the gravitational constant and the permeability of free space respectively and $c_{1}$ is a constant determined numerically by \citet{ms76} to be $c_{1}\approx 0.53$. 

 In this paper, we have performed calculations starting with an initially uniform magnetic field with mass-to-flux ratios in units of the critical value of $M/\Phi = \infty$ (i.e. no magnetic field), 10, 5 and 3. All of our calculations are `supercritical' (that is, unstable to collapse) as under our assumption of ideal MHD (i.e. no ambipolar diffusion or resistivity), subcritical clouds would not (and do not) collapse.

 The corresponding physical field strength for a given mass-to-flux ratio and cloud dimensions is
\begin{equation}
B_{0} = 194~\mu G \left(\frac{M}{\Phi} \right)^{-1} \left(\frac{M}{50~{\rm M}_{\odot}}\right) \left(\frac{R}{0.188~{\rm pc}}\right)^{-2},
\end{equation}
where $M/\Phi$ is the mass to flux ratio in units of the critical value. Thus, a simulation with a critical mass-to-flux ratio would have $B_{0} = 194\mu G$ and for the calculations with mass-to-flux ratios of $\infty, 10, 5$ and $3$ the corresponding field strengths are given by $B_{0} = 0, 19, 39$ and $65 \mu G$, respectively.

 The magnetic field may also be parametrised in terms of the plasma $\beta$, the ratio of gas to magnetic pressure, according to
\begin{equation}
\beta =  0.028 \left(\frac{M}{\Phi} \right)^{2} \left(\frac{c_{s}}{184~{\rm m~s}^{-1}}\right)^{2} \left(\frac{M}{50~{\rm M}_{\odot}}\right)^{-1} \left(\frac{R}{0.188~{\rm pc}}\right).
\end{equation} 
  The simulations presented here thus have initial $\beta$'s of $\infty$, $2.8$, $0.7$ and $0.25$ respectively. Note that the magnetic pressure is dominant over gas pressure in the cloud for mass-to-flux ratios $< 6$ which is the case for the two strongest-field calculations. Indeed, as in \citet{pb08}, we find
that these two calculations show far more significant differences compared to the weaker field and hydrodynamic calculations.

 Finally, the Alfv\'en speed in the initial cloud can be computed using
\begin{equation}
v_{A} =  1.6 \times 10^{5}~{\rm cm~s}^{-1} \left(\frac{M}{\Phi} \right)^{-1} \hspace{-4pt}\left(\frac{M}{50~{\rm M}_{\odot}}\right)^{\frac{1}{2}} \hspace{-4pt} \left(\frac{R}{0.188~{\rm pc}}\right)^{-\frac{1}{2}},
\end{equation}
giving $v_{A} = 0$, $1.6 \times 10^{4}$, $3.1 \times 10^{4}$ and $5.2 \times 10^{4}$ cm~s$^{-1}$ for the calculations in this paper. Thus, the initial turbulent motions in the cloud are super-Alfv\'enic in all cases with Alfv\'enic Mach numbers of $\infty$, $7.3$, $3.8$ and $2.3$, respectively.



\begin{figure*}
   \centering
   \includegraphics[width=0.495\textwidth]{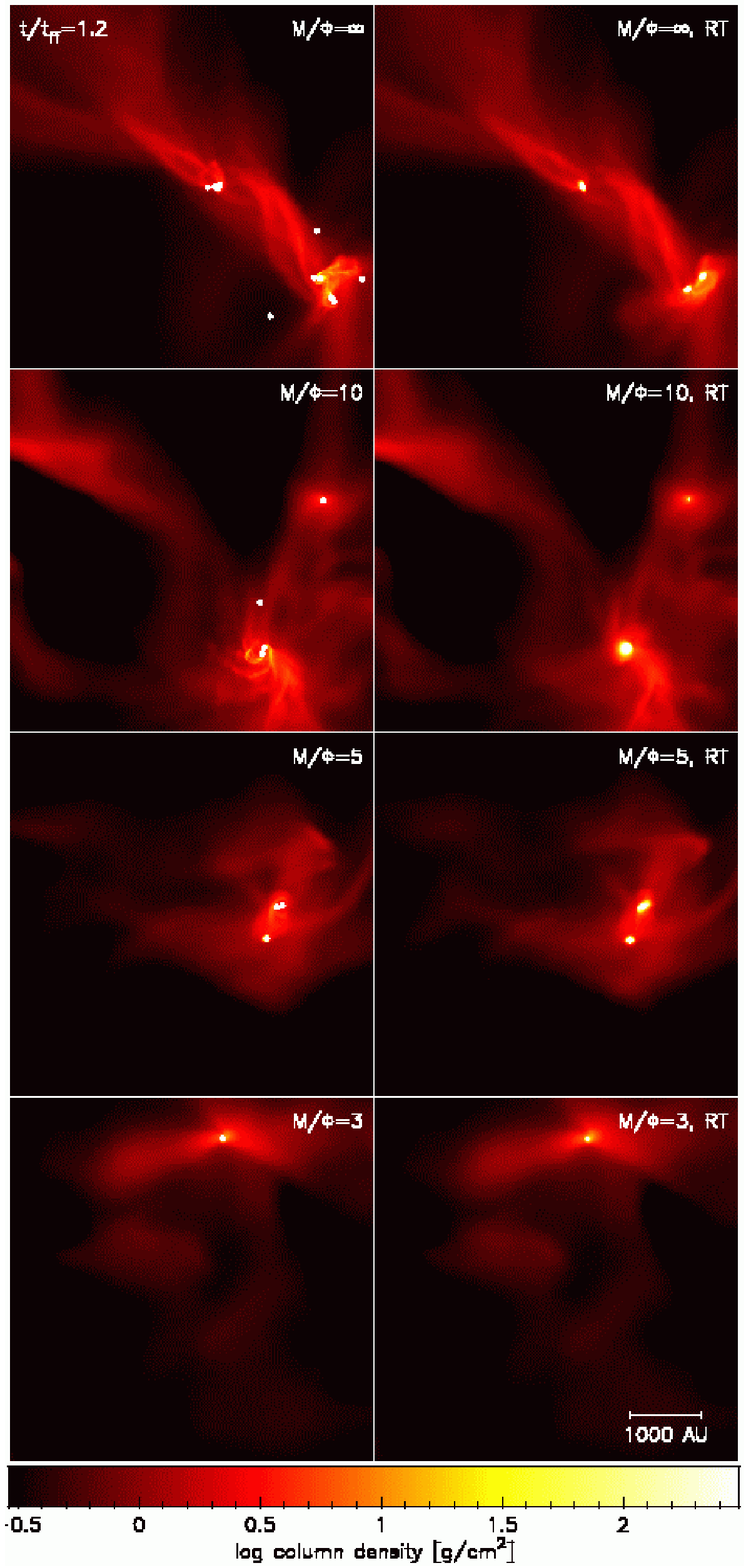}
   \includegraphics[width=0.495\textwidth]{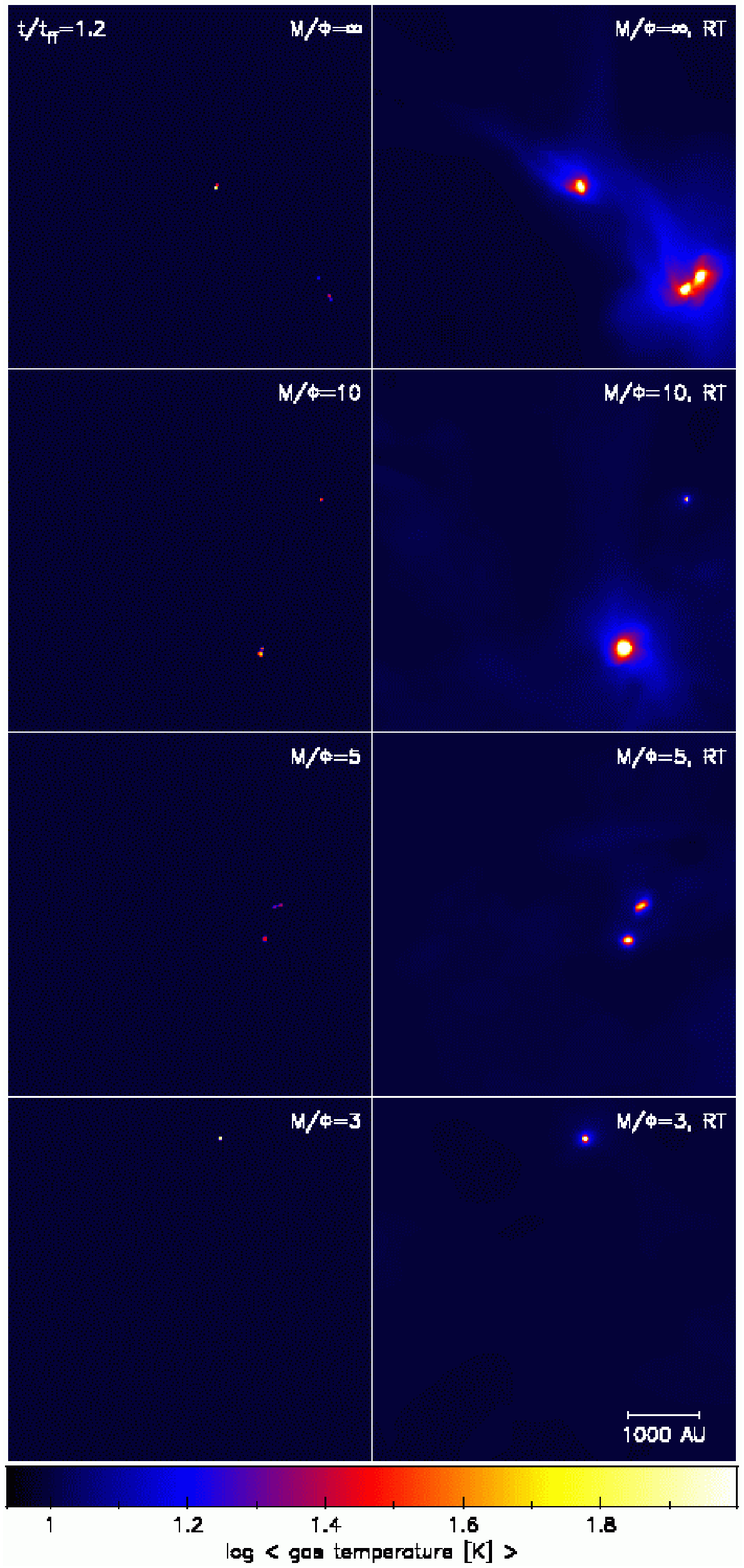}
   \caption{A comparison of the fragmentation that has occurred in the eight different calculations at 1.2 free fall times ($t_{\rm ff}$). The two left-hand columns show column density for each calculation, with magnetic field strength increasing from top to bottom (as indicated by the mass-to-flux ratio in units of the critical value for collapse, where $M/\Phi=\infty$ corresponds to hydrodynamics) using either a barotropic equation of state (first column) or with radiative transfer (second column), as indicated. A strong decrease in protostar formation with increasing magnetic field strength may be observed (comparing rows from top to bottom). The radiative feedback from the protostars is illustrated by plots of the mass-weighted temperature ($\int \rho T~{\rm d}z / \int \rho~{\rm d}z$), shown in the corresponding right-hand panels. The effect of the radiation heating the gas in the vicinity of the protostars (fourth column) can be seen to be poorly captured by the barotropic equation of state approximation (third column) and leads to a dramatic suppression of small-scale fragmentation (comparing the first and second columns).}
   \label{fig:closeup}\label{fig:temp}
\end{figure*}





\section{Results}
\label{sec:results}

 We have computed a total of eight calculations, that is, for four different mass-to-flux ratios, both with and without radiative transfer (where ``without'' means that we use the barotropic equation of state given by equation (\ref{eq:eos}) instead).
 
  The evolution of the simulations can be divided into two stages: i) the initial collapse of the cloud (i.e. up to $\approx 1$ free-fall time) during which the cloud is optically thin, essentially isothermal, and the dynamics and large scale structure are primarily controlled by the interaction of turbulence and magnetic fields; ii) the subsequent evolution of the cloud after the formation of the first star (i.e. $\gtrsim 1$ free-fall time), where the cloud has optically thick regions embedded in the wider (optically thin) large scale structure, and where the small-scale fragmentation is regulated by the radiative feedback from existing protostars on the gas.
 
\subsection{Large-scale cloud structure}
 During the first phase, the radiative transfer has little effect on the overall dynamics compared to the use of a barotropic equation of state because the cloud is optically thin, radiation can escape easily and there are no significant sources of radiation.  Typical temperature variations are of the order $\Delta T/T \sim 10$\%. However, it is computationally very expensive to compute the evolution of the cloud with radiative transfer in the optically thin regime. We have therefore computed only one set of calculations of the global cloud structure during the first (isothermal) phase, the results of which are shown at one free-fall time in Figure~\ref{fig:global} (with magnetic field strength increasing from left to right, as indicated).  These are essentially the same as those presented by \citet{pb08} are we therefore discuss them only briefly here.
 
  Figure~\ref{fig:global}, as in \citet{pb08} reveals the dramatic influence the global magnetic field has on the large scale structure of the cloud, even though the field is much too weak to prevent global gravitational collapse. In particular, for the two strong magnetic field calculations (mass-to-flux ratios of $5$ and $3$ shown in the two right hand panels) large-scale voids are visible in the cloud where material has slipped down the field lines to leave behind evacuated but magnetically-pressurised voids. These magnetic-pressure supported voids were discussed in detail in \citet{pb08} (see also \citealt{pbd08}) and appear in the regime where $\beta < 1$ (i.e. where the magnetic pressure is dominant over the gas pressure). This regime is particularly interesting given that almost all magnetic field strength measurements in molecular clouds indicate that $\beta < 1$ \citep{crutcher99,bourkeetal01,ht04,hc05}. Also visible during the initial expansion phase is a `stripy' structure in the column density maps which is aligned with the large scale magnetic field lines.  This is a consequence of the anisotropy of turbulent motions in the presence of a magnetic field \citep[e.g.][]{gs95} and, while not so obvious in Figure~\ref{fig:global}, was discussed and clearly illustrated by \citet{pb08}.

\subsection{Fragmentation}
 From one free-fall time ($t_{\rm ff}$), the cloud structures shown in Figure~\ref{fig:global} were evolved both with and without radiative transfer (i.e. using the barotropic equation of state in the former case and the full flux-limited diffusion equations in the latter). The simulations were run from this point to between 1.25 and $1.54~t_{\rm ff}$ ($2.93 \times 10^5$ yrs) depending on the computational expense (the calculations slow down significantly once star formation initiates and the more protostars are formed, the slower the calculations become). The barotropic calculations with mass-to-flux ratios of $M/\Phi=\infty, 10, 5$ and 3 begin forming stars at $t \approx 1.07, 1.03, 1.10$ and $1.19 t_{\rm ff}$, respectively, with the star formation in the radiative transfer conterparts typically being delayed by $\approx 0.01~t_{\rm ff}$.  A close-up of the fragmentation in all eight simulations is shown in Figure~\ref{fig:closeup}, showing column density (left-hand panels) and mass-weighted temperature (right-hand panels) at $1.20~t_{\rm ff}$, after star formation has begun in all eight clouds.   The sink particles are shown as white filled circles.
 
  The left-hand (column density) panels of Figure~\ref{fig:closeup} dramatically illustrates two main effects. The first is an overall decrease in star formation rate with increasing magnetic field strength (rows from top to bottom are in order of weakest to strongest magnetic field).  This is a result of the influence of the global magnetic field on the large scale cloud structure, as already evident in Figure~\ref{fig:global}. In particular, for the stronger field calculations (bottom two rows of Figure~\ref{fig:closeup}, and the rightmost two panels of Figure~\ref{fig:global}), large parts of the cloud are supported against collapse by the magnetic field resulting in fewer collapsing sub-regions (or `cores'). For example, where the hydrodynamic calculation (top row of Figure~\ref{fig:closeup}) has collections of protostars separated by a couple of thousand AU, only one collapsing region is evident in the strongest field ($M/\Phi=3$) case (bottom row), which shows no sub-fragmentation either with or without radiative transfer. The effect of the magnetic field in slowing the infall from the global cloud is further quantified in Figure~\ref{fig:sinkmass} and discussed in Section \ref{sec:sfr}, below.
  
  The second effect visible in Figure~\ref{fig:closeup} is the dramatic suppression of small-scale fragmentation by the radiative feedback. This is especially obvious in the hydrodynamic/weak field calculations where the calculations using a barotropic equation of state have fragmented into multiple low-mass objects which interact violently, causing ejections of very low mass objects from multiple systems. This small-scale fragmentation occurs primarily in the massive protostellar discs.  By contrast, in the calculations which include radiative transfer, all of the subsequent disc fragmentation is suppressed by the radiation from the existing protostar(s). A good example is found in the lower-right of the $M/\Phi=10$ panels of Figure~\ref{fig:closeup}: with radiative feedback a single object with a disc is formed, while without radiative feedback this disc fragments into three objects, one of which is ejected. The radiative feedback, in effect sets a minimum distance between protostars by substantially increasing the temperature and therefore the Jeans length in the gas immediately surrounding a protostar \citep{bate09b}.

 The differences between computing the radiative transfer and using the barotropic equation of state approximation are best illustrated by plotting the temperature, given in the right-hand panels of Figure~\ref{fig:temp}. Each panel shows the integrated temperature map (i.e. $\int \rho T ~{\rm d}z / \int \rho ~{\rm d}z$) for the corresponding column-density panel in Figure~\ref{fig:closeup}.  We have not plotted the sink particles on these panels so that the temperature distribution very close to the protostars can be seen for the barotropic calculations. For the barotropic equation of state (centre-right panels), the temperature is simply related to the density, leading to very point-like sources of energy concentrated around the protostars themselves. In the radiative transfer calculations (right-most panels), the radiation emitted from the high-density optically-thick gas near the protostars heats a much larger surrounding region to temperatures $> 30$~K, effectively shutting off any further fragmentation in this material (as evident in the left-hand column-density panels of Figure~\ref{fig:closeup}).

\begin{figure}
   \centering
   \includegraphics[width=\columnwidth]{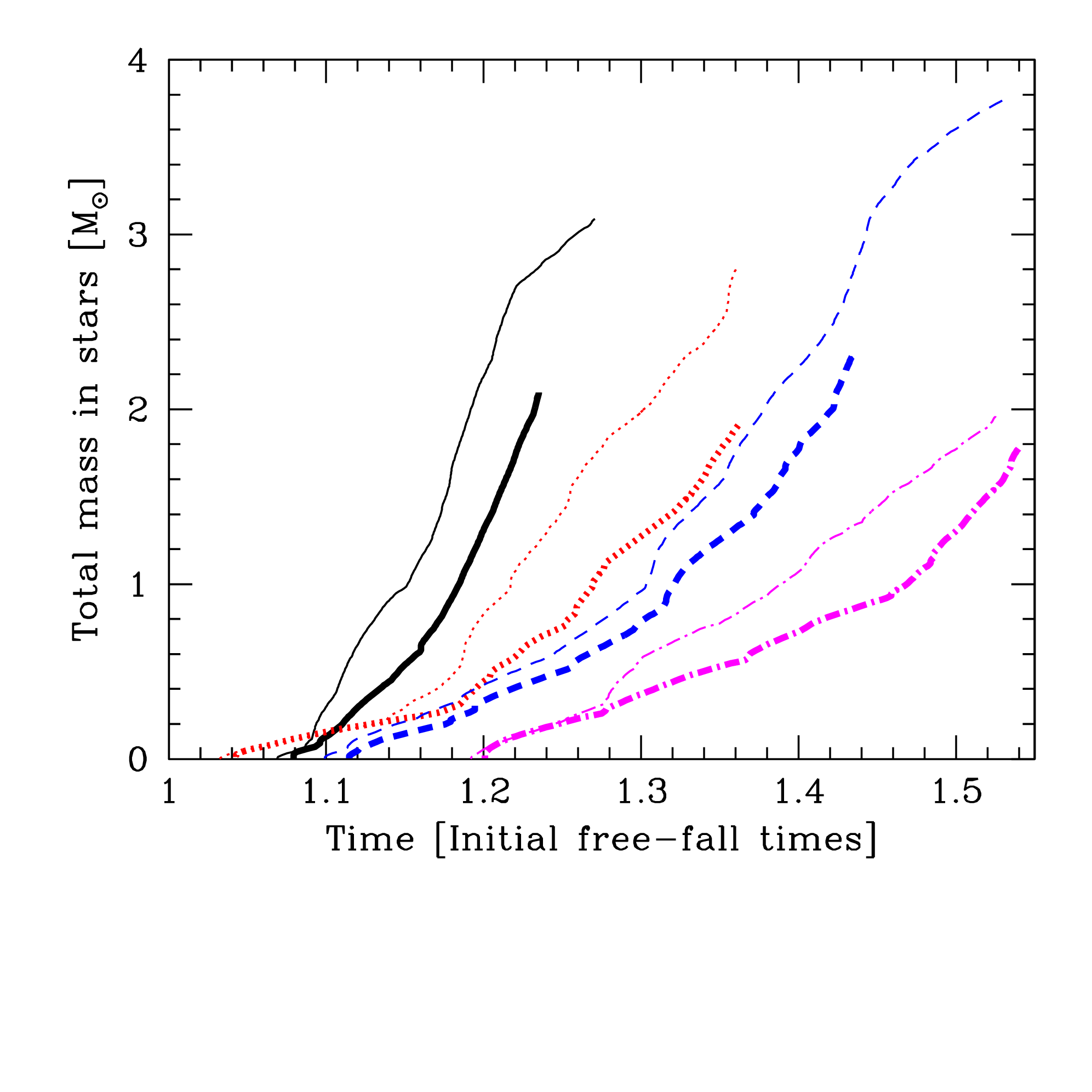} 
   \caption{Total mass in stars (sink particles) as a function of time, showing all eight calculations with (thick lines) and without (thin lines) radiative transfer at four different magnetic field strengths: hydrodynamic (solid black lines), $M/\Phi=10$ (dotted red lines), $M/\Phi=5$ (dashed blue lines), and $M/\Phi=3$ (dot-dashed magenta lines).  The star formation rate decreases with increasing magnetic field strength and with the addition of radiative feedback.  Note how the two curves for each magnetic field strength track each other for some time before diverging, indicating that radiative feedback only plays a role in suppressing subsequent fragmentation rather than changing the initial pattern of star formation.}
   \label{fig:sinkmass}
\end{figure}

\begin{figure}
   \centering
   \includegraphics[width=\columnwidth]{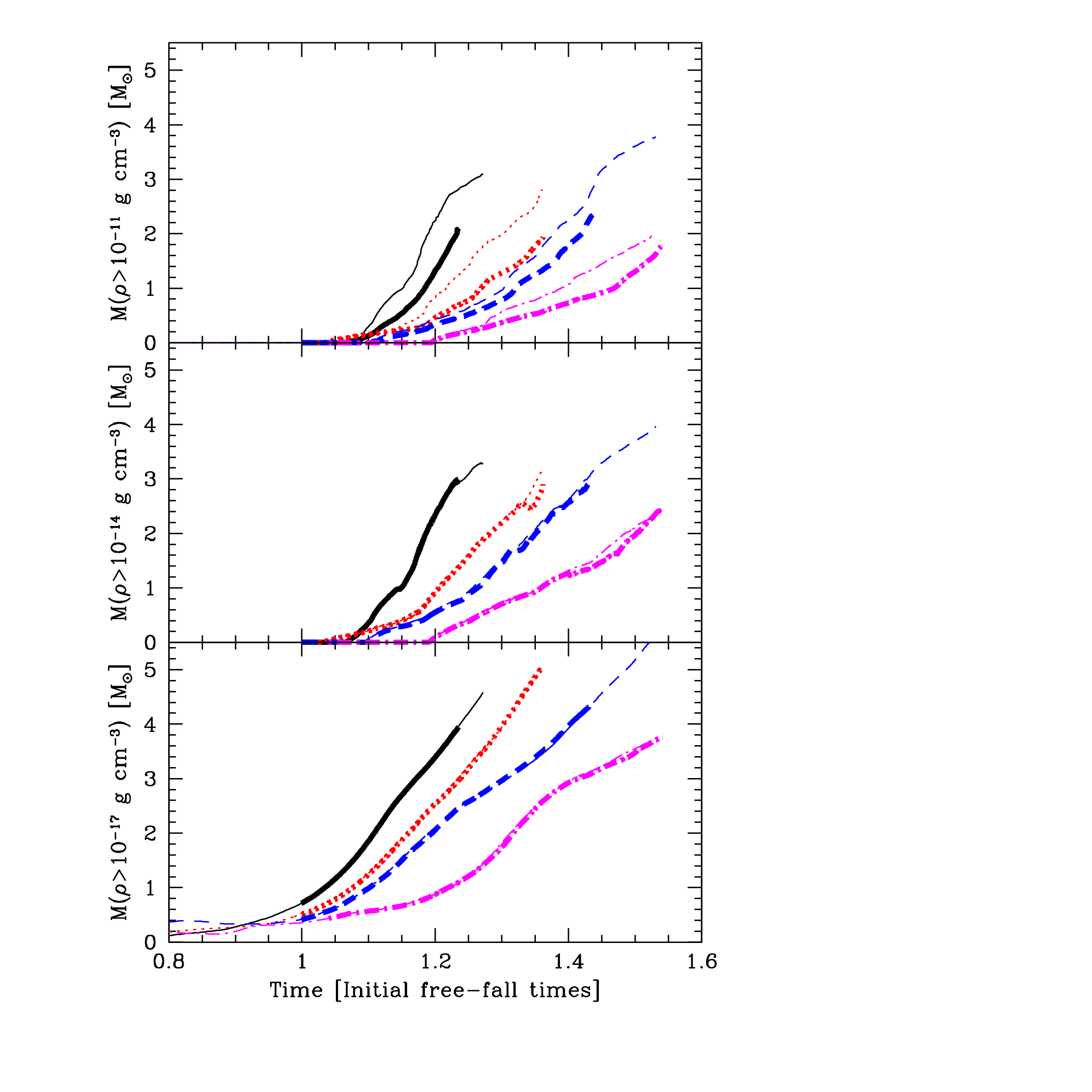} 
   \caption{The total mass above certain density thresholds in each collapsing cloud as a function of time. From top to bottom the panels show $M (\rho > 10^{-17}$g~cm$^{-3})$ (approximately two orders of magnitude denser than the original cloud density), $M (\rho > 10^{-14}$g~cm$^{-3})$, and $M (\rho > 10^{-11}$g~cm$^{-3})$ (i.e. above which most material is in protostars).  The different lines are as in Figure \ref{fig:sinkmass}.  Thick lines denote those calculations with radiative feedback, while thin lines are using the barotropic equation of state.  The line types and colours denote the magnetic field strength (also ordered from top to bottom in each panel with progressively increasing magnetic field strength). Magnetic fields can be seen to affect the collapse rate at all density thresholds (all panels), while radiative feedback primarily prevents fragmentation in the highest density regions of the cloud (top panel, comparing thin and thick lines).}
   \label{fig:massaboverho}
\end{figure}

\subsection{Star formation rate}
\label{sec:sfr}
 The effects of both magnetic fields and the radiative feedback on the star formation rate are quantified in Figure~\ref{fig:sinkmass}, which shows the total mass in protostars (that is, the total mass of all sink particles in a simulation) as a function of time. After $t=1.2~t_{\rm ff}$, the eight simulations form a strict sequence of progressively decreasing star formation rate in the order: hydrodynamic, barotropic; hydrodynamic, RT; $M/\Phi=10$, barotropic,  $M/\Phi=10$, RT; $M/\Phi=5$, barotropic,  $M/\Phi=5$, RT; $M/\Phi=3$, barotropic,  $M/\Phi=3$, RT, i.e. with magnetic fields as the primary effect and radiative feedback secondary. The rate at which gas is converted into stars decreases due to the influence of both magnetic fields and radiative feedback, though more strongly with the former. For example, at $t=1.2~t_{\rm ff}$, the hydrodynamic calculation contains $2.2 ~{\rm M}_{\odot}$ in stars using a barotropic equation of state, compared to $1.3 ~{\rm M}_{\odot}$ with radiative feedback, both of which are higher than the $[0.83 ~{\rm M}_{\odot}, 0.44 ~{\rm M}_{\odot}]$ formed by the weak field $M/\Phi=10$ calculation at the same time [without,with] radiative transfer. These numbers decrease further to $[0.42 ~{\rm M}_{\odot}, 0.33 ~{\rm M}_{\odot}]$ for the $M/\Phi = 5$ simulation and further still to $[0.055 ~{\rm M}_{\odot}, 0.0056 ~{\rm M}_{\odot}]$ for the strongest magnetic field case ($M/\Phi=3$). This general trend is continued as far as we have been able to run the calculations in each case (Figure~\ref{fig:sinkmass}).
 
 The fact that the radiative feedback influences \emph{subsequent} star formation rather than the initial fragmentation is also evident from Figure~\ref{fig:sinkmass}. In particular, the two curves corresponding to the same magnetic field strength but with and without radiative transfer in each case track each other closely after first sink formation, before diverging at later times. Taking the $M/\Phi=3$ case as an example (i.e. the lower two curves in Figure~\ref{fig:sinkmass}), and comparing the time evolution in Figure~\ref{fig:sinkmass} to the fragmentation sequence shown in Figure~\ref{fig:time}, it may be observed that the two curves diverge when secondary disc fragmentation occurs in the barotropic calculation ($t\approx 1.27~t_{\rm ff}$), leading to a burst of star formation (and subsequent ejection of low-mass objects from the multiple system). In the radiative transfer case, the disc does not fragment but instead continues to slowly accrete onto the existing protostar.

 Magnetic fields and radiation are also found to affect different densities in the cloud. Figure~\ref{fig:massaboverho} shows the mass above a given density threshold in the cloud as a function of time for three different density thresholds, $\rho > 10^{-17} {\rm g~cm}^{-3}$ (bottom panel), $\rho > 10^{-14}{\rm g~cm}^{-3}$ (middle panel), and $\rho > 10^{-11} {\rm g~cm}^{-3}$ (top panel), where solid lines correspond to calculations using a barotropic EOS and dashed lines refer to calculations using radiative transfer and, as in Figure \ref{fig:sinkmass}, the lines form a sequence from top to bottom with increasing magnetic field strength. At a density threshold of $10^{-17}{\rm g~cm}^{-3}$ (bottom panel), whilst there is a strong decrease in the mass collapsing to higher densities with increasing magnetic field strength, there is almost no difference between the barotropic simulations and those with full radiative transfer (i.e. comparing the solid and dashed lines), indicating that radiative feedback plays very little role at these densities. At a threshold of $\rho > 10^{-14}{\rm g~cm}^{-3}$ (middle panel) the results are similar (although the overall masses are lower) apart from some divergence at $t \gtrsim 1.4 t_{\rm ff}$ in the $M/\Phi=3$ calculation. By contrast, at higher densities ($\rho > 10^{-11}{\rm g~cm}^{-3}$, top panel), where the gas is optically-thick to radiation, there are differences of up to $\sim 50\%$ in the mass above this density between the barotropic and radiative transfer calculations (the latter having systematically lower mass accumulation rates) similar to the differences observed in Figure~\ref{fig:sinkmass}.

\begin{figure*}
   \centering
   \includegraphics[width=0.45\textwidth]{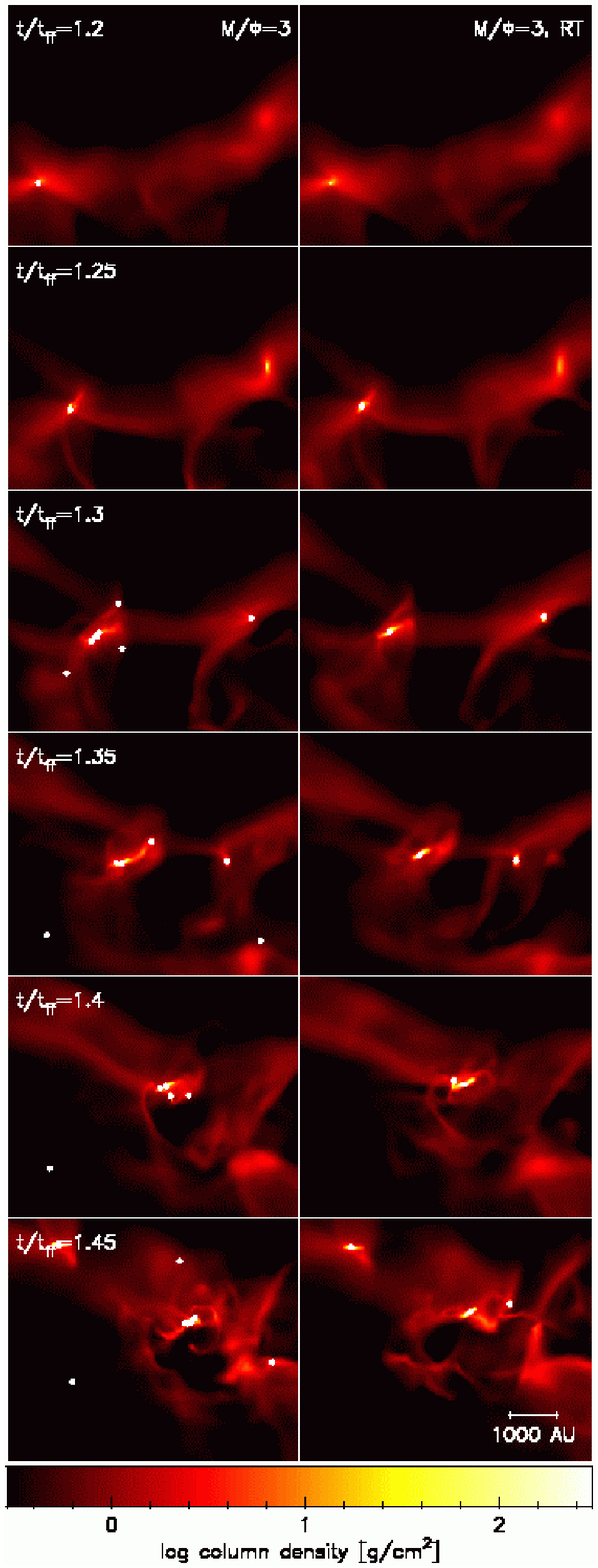}
   \includegraphics[width=0.45\textwidth]{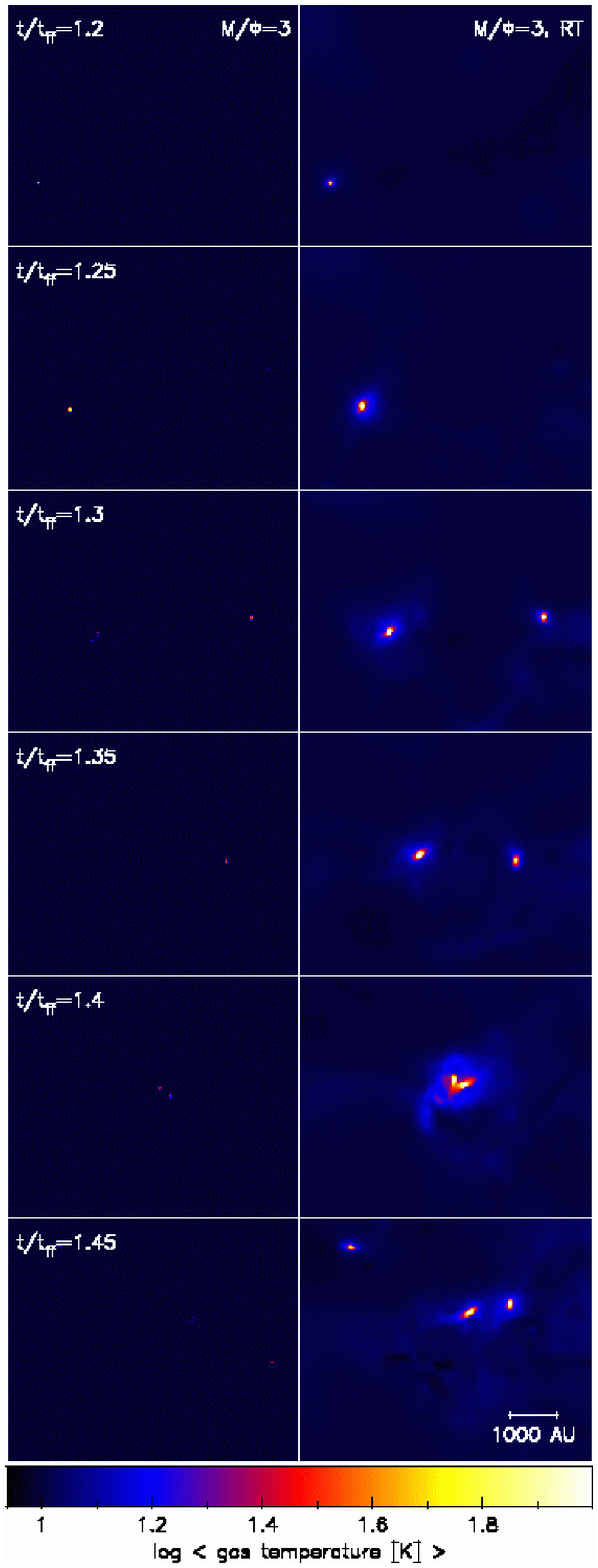}
  \caption{Time sequence of fragmentation in the strongest magnetic field calculation ($M/\Phi = 3$), shown from the comparison time of $t = 1.2~t_{\rm ff}$ in Figure~\ref{fig:closeup} (top row, where the panels here are shifted in position relative to Figure~\ref{fig:closeup} to follow the subsequent fragmentation) up to $t = 1.45_{\rm ff}$ (bottom row) at intervals of $0.05 t_{\rm ff}$. The two left-hand columns show column density for the calculation using a barotropic equation of state (first column) and with radiative transfer (second column). Corresponding mass-weighted temperatures for the two calculations are shown in the two right-hand columns, highlighting the heating of the gas due to the radiative feedback in the regions immediately surrounding the protostars. Although the radiative feedback suppresses fragmentation on the smallest scales, dynamical interactions nevertheless occur over larger length scales, as evident from the merger of the two star-forming cores which occurs at $t = 1.4 t_{\rm ff}$ in each of the calculations.}
   \label{fig:time}
\end{figure*}

\subsection{Dynamics}
 The effect of the reduced fragmentation on the dynamics of the protostars due to the radiative feedback is illustrated in Figure~\ref{fig:time}, showing a time sequence of the evolution in the strongest magnetic field case ($M/\Phi = 3$) in intervals of $0.05~t_{\rm ff}$ (9,500 yrs) from the onset of star formation. As in Figure~\ref{fig:closeup}, the two left-hand columns show column densities for the barotropic (first column) and radiative transfer (second column) simulations, whilst the corresponding right-hand panels show the mass-weighted temperature. The overall picture is similar to that apparent from Figure~\ref{fig:closeup}, except that the time sequence shows that despite the fact that the $M/\Phi=3$ simulations each only produced a single object at $t=1.2~t_{\rm ff}$ collapse and fragmentation continues to form binary and multiple systems at later times (i.e. $t \gtrsim 1.3~t_{ff}$). However, whilst the larger scale dynamics is similar between the barotropic and RT calculations (e.g. the merger of the two main collapsed regions at $t \approx 1.4~t_{\rm ff}$), the smaller scale disc fragmentation which results in a triple ejection at $t=1.3 t_{ff}$ in the barotropic calculation is completely absent from the radiative transfer simulation, as a result of the higher temperatures surrounding the first protostars to form (fourth column of Figure~\ref{fig:time}).

The combined effects of stronger magnetic fields and radiative feedback are, therefore, to decrease the star formation rate (Figure \ref{fig:sinkmass}) and decrease the number of occurrences of dynamical interactions and ejections between protostars.

\begin{table*}
\begin{tabular}{|cc|cccc|ccccc|}\hline
\multicolumn{2}{|l|}{Calculation} & \multicolumn{4}{c|}{Time: 1.25 $t_{\rm ff}$} & \multicolumn{5}{c|}{End of Calculation} \\ \hline
Barotropic or & $M/\Phi$ & Number & Mass & Mean Mass & Median Mass & Time & Number & Mass & Mean Mass & Median Mass \\
Radiative Transfer & & & M$_\odot$ & M$_\odot$ & M$_\odot$ & $t_{\rm ff}$ & & M$_\odot$ & M$_\odot$ & M$_\odot$\\ \hline
Barotropic & $\infty$ & 17 & 2.93 & 0.17 & 0.11 & 1.274 & 17  & 3.11 & 0.18 & 0.13 \\
 & 10 & 10 & 1.41& 0.14 & 0.04 & 1.361 & 21 & 2.82 & 0.13 & 0.06 \\
 & 5 & 6 & 0.64 & 0.11 & 0.12 & 1.531 & 23  & 3.77 & 0.16 & 0.12 \\
 & 3 & 1 & 0.22 & 0.22 & 0.22 & 1.525 & 18  & 1.96 & 0.11 & 0.06  \\
RT &  $\infty$ & -- & -- & -- & -- & 1.235 & 10  & 2.09 & 0.70 & 0.78 \\
& 10 & 2 & 0.75 & 0.38 & 0.38 & 1.362 & 5 & 1.92 & 0.38 & 0.14 \\
& 5 & 4 & 0.50 & 0.13 & 0.13 & 1.437 & 10  & 2.34 & 0.23 & 0.21 \\
& 3 & 1 & 0.21 & 0.21 & 0.21 & 1.541 & 7  & 1.80 & 0.26 & 0.18 \\ \hline
\end{tabular}
\caption{The statistical properties of the protostars formed in the eight calculations.  For each of the four mass-to-flux ratios, barotropic and radiative transfer calculations were performed.  Due to computational expense, the calculations were followed for a different amounts of time.  All but one calculation was evolved until 1.25~$t_{\rm ff}$, so we give the statistical properties of the simulations at this time.  We also give the statistical properties at the end of each calculation.  In each case, we give the number of protostars (sink particles) formed, the total mass in protostars, and the mean and median masses of the protostars.  It is clear that using a barotropic equation of state produces many more objects than are obtained with radiative transfer.  It is also clear that the rate of protostar production decreases strongly with magnetic field strength in the barotropic calculations (column 3), although the typical mass of the protostars is independent of field strength.  Conversely, with radiative transfer there is no significant dependence of the rate of protostar production with magnetic field strength, and there is an indication that the mean masses of the protostars may decrease with decreasing field strength (columns 5 and 10). }
\label{table:mass}
\end{table*}

\subsection{Protostellar masses}

The calculations presented here are of 50~M$_\odot$ clouds that collapse to form 3--23 protostars.  With such small numbers of objects, and the fact that the simulations have not all been followed for the same amount of time, it does not make sense to attempt to plot stellar mass functions.  Rather, in Table \ref{table:mass}, we give the amount of mass that has been converted into protostars (sink particles), the number of protostars, and the mean and median masses of the protostars.  We give these values at $t=1.25~t_{\rm ff}$ for all but one of the calculations, and at the end of each calculation.

Generally, as found by \citet{bate09b}, the effect of radiative feedback is to dramatically decrease the number of protostars formed compared with the barotropic equation of state (Table \ref{table:mass}, columns 3 and 8).  Simultaneously, the protostars are generally found to be more massive with radiative feedback because gas that would have formed other objects via the fragmentation of discs and nearby filaments using a barotropic equation of state is hotter and is able to be accreted by existing protostars instead (Table \ref{table:mass}, column 5 for $M/\Phi=5,10$ and column 8 for $M/\Phi=\infty$).  In the strongest magnetic field case, these statements are still true, but the trends only become apparent fairly late in the calculations because of the delay of the star formation caused by the strong field (Table \ref{table:mass}, column 8 for $M/\Phi=3$).

When investigating the effect of the magnetic field things become more interesting.  As already discussed, the rate at which gas is converted into stars decreases strongly with increasing magnetic field strength for both the barotropic and radiative transfer calculations  (Table \ref{table:mass}, column 4).  However, where this mass goes differs significantly between the barotropic and radiative transfer calculations.  For the barotropic calculations, the rate of protostar formation decreases strongly with increasing magnetic field strength, but the typical masses of objects are independent of the magnetic field strength  (Table \ref{table:mass}, columns 5, 6, 10, and 11).  We also note that if the calculations are followed for a long periods of time all of the calculations eventually produce large numbers of objects regardless of the field strength (Table \ref{table:mass}, column 8).  However, with radiative feedback there is no significant dependence of the rate of protostar formation on the magnetic field strength (Table \ref{table:mass}, columns 3 and 8), and there is an indication that the mean masses of the protostars may increase with decreasing magnetic field strength (Table \ref{table:mass}, columns 5 and 10).  This latter effect is presumably because more of the gas is supported with a stronger magnetic field and not able to be accreted by the protostars.  Although this needs to be confirmed with larger calculations that form larger numbers of objects, this implies that the characteristic stellar mass may decrease with increasing magnetic field strength, a result that is somewhat counterintuitive since a naive calculation of a magnetic Jeans mass would lead one to conclude that the characteristic stellar mass should increase with increasing magnetic field strength.

\section{Discussion}
\label{sec:discussion}

 In this paper we have studied, for the first time, the combined effects of magnetic fields and radiative feedback on the formation of stellar clusters from turbulent molecular clouds. We find that the two effects are complementary in the sense that they affect the star formation process at very different scales. Magnetic fields affect the large-scale cloud structure (Figure~\ref{fig:global}), influencing all densities in the cloud (Figure~\ref{fig:massaboverho}).  Stronger fields decrease the overall star formation rate (Figure~\ref{fig:sinkmass}). By contrast, radiative feedback affects small-scale fragmentation (Figure~\ref{fig:closeup}) and influences only the highest densities in the cloud (Figure~\ref{fig:massaboverho}).  It influences the star formation rate primarily by inhibiting small-scale fragmentation in cores once the first protostar has been formed (Figures~\ref{fig:closeup}, \ref{fig:sinkmass} and \ref{fig:time}). However, multiple systems are still common, formed from well-separated but mutually bound condensations (Figure~\ref{fig:time}).

 The primary effect of the magnetic field is to lower the accretion rate onto the star-forming cores by providing large-scale support to low-density regions of the cloud, thus preventing this material from subsequently being accreted. There is no clear shift in the \emph{onset} of star formation with magnetic field strength (Figure~\ref{fig:sinkmass}), except perhaps in the strongest magnetic calculation where star formation (ie. sink particle creation) does not initiate until $t\approx 1.20 t_{\rm ff}$ (Figures~\ref{fig:time}~and~\ref{fig:sinkmass}) compared to $t\approx 1.03-1.11 t_{\rm ff}$ in the moderate/weak/zero field simulations.  Since the simulations do not produce large numbers of protostars and they are not followed very far any conclusions regarding the masses of the protostars must be treated with caution.  However, we find that using a barotropic equation of state the typical masses of the protostars do not depend significantly on the magnetic field strength but the number of protostars formed increases with weaker fields (Table \ref{table:mass}).  Conversely, with radiative feedback, the numbers of protostars formed in the clouds does not vary greatly with the magnetic field strength but the masses of the protostars tend to be lower with stronger magnetic fields.  Generally, radiative feedback results in a larger characteristic protostellar mass than using the barotropic approximation.

 The general effect of radiation on the fragmentation is easily understood in terms of the increase in the Jeans length of the heated gas surrounding existing protostars. An increase in temperature (e.g. from $10K$ to $\gtrsim 30K$ as in Figure~\ref{fig:temp}) leads to an increase in the Jeans length since $\lambda_{J} \propto T^{1/2}$. Because radiative feedback acts mainly on small scales, it takes longer for this to affect the overall star formation rate substantially (e.g. note the reduction in the figures of $M(>\rho)$ propagating slowly to higher densities in the $M/\Phi=3$ run in Figure~\ref{fig:massaboverho} due to the progressive heating of wider regions of the cloud visible in Figure~\ref{fig:time}).  However, it has a dramatic influence on the initial mass function (IMF) by suppressing fragmentation in discs (and nearby filaments) and decreasing the likelihood of forming multiple systems from which low-mass members can be ejected (e.g. Figure~\ref{fig:time}).  The effect of the radiative feedback is also more pronounced when the potential well in which the protostars form is deeper, partially offsetting the inherent decrease of the Jeans mass with increasing density, which in turn leads to a reduced dependence of the IMF on the initial density of the cloud which, as discussed in detail by \citet{bate09b}, may explain why the IMF appears to be so universal across very different star-forming environments.

\begin{figure*}
   \centering
   \includegraphics[angle=270,width=0.48\textwidth]{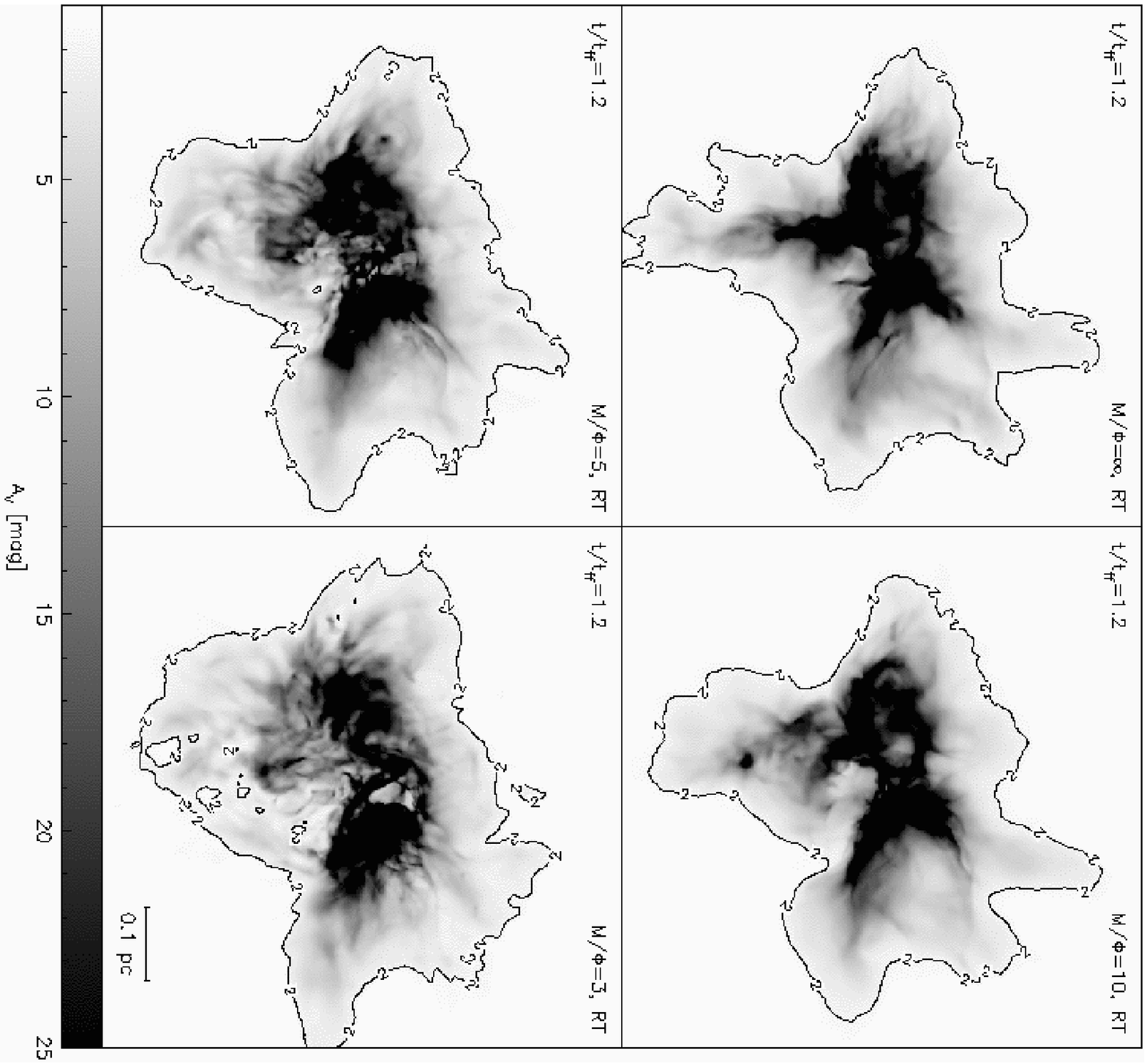}\hspace{0.0\textwidth}
   \includegraphics[angle=270,width=0.48\textwidth]{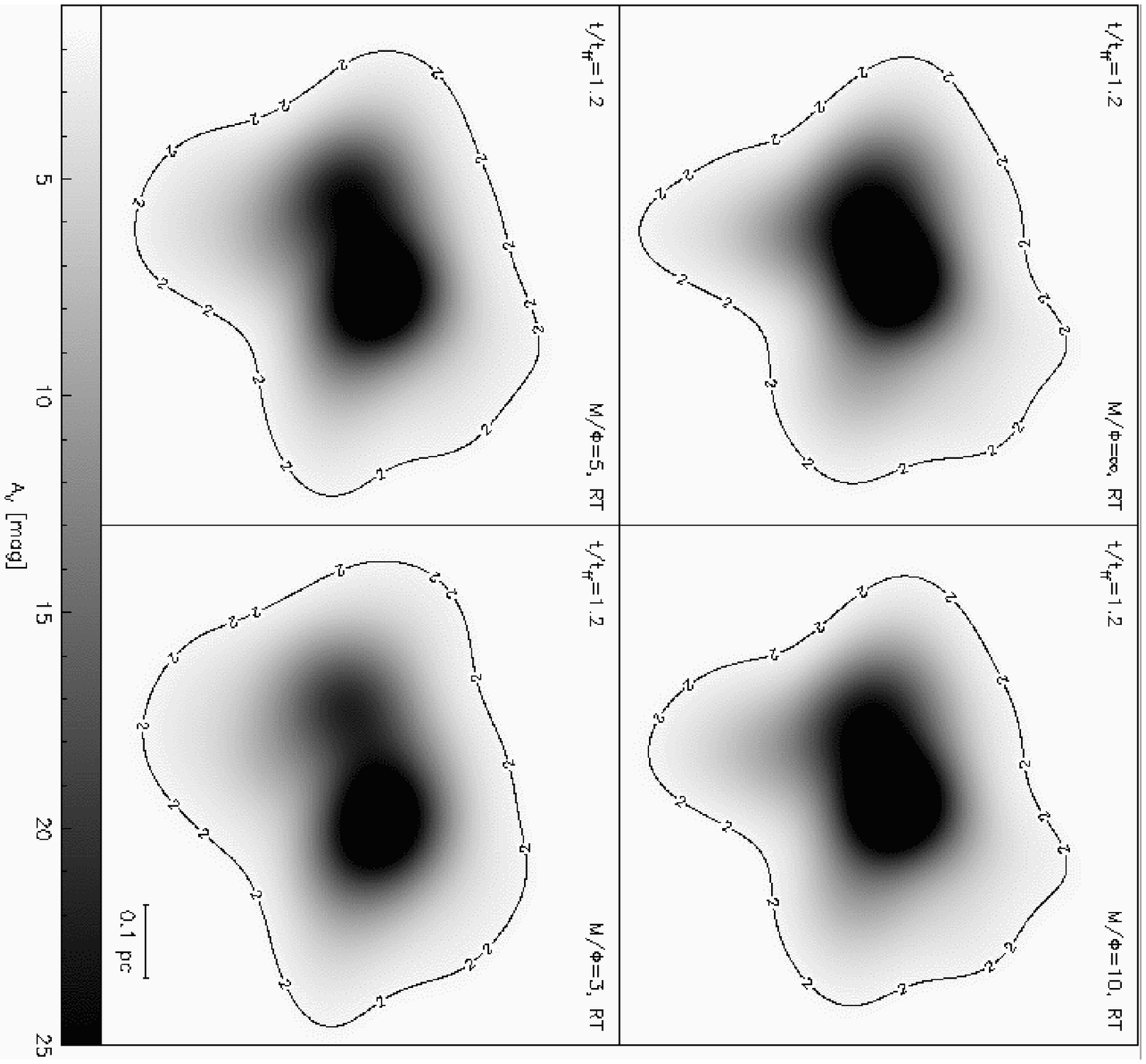}
  \caption{Extinction maps of the four calculations with radiative transfer and varying initial magnetic field strength ($M/\Phi$ as indicated) at the simulation resolution (left) and at the resolution of the \citet{evansetal09} extinction maps for Ophiuchus (right). The $A_{V}=2$ contour is shown, the mass inside of which was used to calculate the mass in the \emph{c2d} survey. We have converted hydrogen column density to $A_{V}$ using the inverse of the conversion factor adopted for the \emph{c2d} observations by \citet{evansetal09}, i.e. a conversion from extinction to hydrogen column density of $1.37\times 10^{21} {\rm cm}^{-2} {\rm mag}^{-1}$. The scale from $A_{V}=1$ to $A_{V}=25$ is the same as that used for the \emph{c2d} maps.}
\label{fig:extinction}
\end{figure*}

 \subsection{Comparison with observations}
 
 With a cloud of only 50 solar masses it is difficult to make a statistically meaningful comparison with observed star-forming molecular clouds as a whole since nearby clouds typically contain $10^{3}-10^{5} ~{\rm M}_{\odot}$ of material over areas as large as $74$ pc$^{2}$ \citep{evansetal09}. Rather our simulated clouds fall within the definition of a `millimeter core', ie. $n_{{\rm H}_{2}} \gtrsim 2 \times 10^{4}$ cm$^{-3}$, and sizes similar to the typically measured core sizes of $1.5 \times 10^{4}$ AU in Ophiuchus and $3 \times 10^{4}$AU in Perseus and Serpens \citep{enochetal07}. As pointed out by \cite{bbb03}, the dense cores formed in simulations of the size presented here are similar to the Ophiuchus-F core which measures $\approx 0.1$~pc across and has a mass of $\approx 8$~M$_\odot$ \citep*{motteetal98}. In Figure~\ref{fig:extinction} we show simulated extinction maps of the four runs with radiative feedback, at the simulation resolution (left) and at the resolution of the \citet{evansetal09} extinction maps for Ophiuchus (right), with extinction on a linear greyscale map from $A_{V}=1$ to $A_{V}=25$ which may be directly compared to the \citet{evansetal09} maps. To produce the extinction maps we have simply used the inverse of the conversion from extinction to hydrogen column density of $1.37\times 10^{21} {\rm cm}^{-2} {\rm mag}^{-1}$ adopted by \citet{evansetal09}. The resolution of the right hand panels is calculated from the \emph{c2d} extinction map resolution of $270''$, which at the assumed distance of $125$pc for Ophiuchus gives a resolution of $0.16$ pc. We have simulated this resolution in our maps by enforcing a minimum smoothing length of $0.08$pc on the SPH particles when calculating the column density (i.e., approximating the point spread function for the extinction maps by the SPH kernel smoothing function).

 Star formation efficiencies are calculated by \citet{evansetal09} by dividing the mass in Young Stellar Objects (YSOs, defined as objects with infrared excesses assumed to correspond to the presence of a disc) by the total mass of the cloud plus YSOs. That is,
\begin{equation}
SFE = \frac{M_{*}}{M_{*} + M(cloud)},
\label{eq:sfe1}
\end{equation}
  where $M(cloud)$ is derived by integrating the extinction maps, converted to column density, over area. Despite the low resolution of the observations compared to our simulated cloud, the cloud masses measured from the clouds on the right hand side of Figure~\ref{fig:extinction}, by integrating column density over the area within the $A_{V}=2$ contour, are remarkably accurate. For example, the measured mass for the zero magnetic fields case ($M/\Phi = \infty$) at the observational resolution is $44.1$ M$_{\odot}$, which may be compared with the total cloud mass in our simulations of 50 M$_{\odot}$, of which $\sim44$ M$_{\odot}$ lies within the sphere with the approximate radius of the $A_{V}=2$ contour. The caveat to this for the observations is that the conversion from extinction to column density relies on a model for the dust, changes to which can have a significant impact on cloud masses (e.g. \citealt{evansetal09} discuss the fact that their cloud masses are revised down by a factor of $1.4$ from previous estimates due to revision of the dust model).

Efficiencies thus derived by \citet{evansetal09} range from $3-6\%$, which is assumed to represent an average over the last 2~Myr given that this is the estimated lifetime of YSOs with infrared excesses.
The comparison between observations and our simulations is made more difficult by the fact that we are not able to follow the cloud collapse for longer than around $1.5 t_{\rm ff}$ with current computational resources\footnote{The key limitation being that for only a few collapsed objects, good load balancing is very difficult to achieve, limiting the usefulness of simply running on a higher number of processors.}. Nevertheless, one can make tentative estimates based on the star formation rates we find in Figure~\ref{fig:sinkmass} and the masses in Table~\ref{table:mass}. For the four clouds shown on the right hand side of Figure~\ref{fig:extinction} (the four runs with radiative feedback included), cloud masses measured by integrating the column density within the $A_{V}=2$ contour are $44.1$, $44.0$, $43.6$ and $43.9$ for the $M/\Phi=\infty$, $10$, $5$ and $3$ clouds respectively. A straightforward application of (\ref{eq:sfe1}) at the end point of each of our calculations indicates that of order $3-5\%$ of the gas has been converted into stars over the time for which the simulations have been run. Whilst these values are in agreement with the observational results (though not for dense gas), they are not very meaningful given that they represent evolution over fractions of a freefall time beyond initial star formation (the end time for each of the calculations is given in Table~\ref{table:mass} and can be inferred from Figure~\ref{fig:sinkmass}) and will increase with time as more mass is converted into stars.

 More useful are estimates which take into account the timescale over which star formation has proceeded. The depletion time for the cloud is given by
\begin{equation}
t_{dep} = M(cloud)/\dot{M}_{*}.
\end{equation}
 Calculating the average star formation rate from the onset of star formation using Figure~\ref{fig:sinkmass} and assuming that these rates will continue indefinitely (a dubious assumption), we obtain depletion times of $0.8$, $1.6$, $1.3$ and $1.8$~Myr for the clouds in the above four calculations respectively. Whilst these are very short compared to the global depletion times for the clouds in \citet{evansetal09} of $30$ to $66$ Myr, they are in agreement with the depletion timescale derived for dense cores within such clouds (i.e., gas with $n \gtrsim 2 \times 10^{4}$cm$^{-3}$, which our initial cloud density lies above) which are in the range of $0.6$-$2.9$~Myr with an average of $1.8$Myr.

 Finally, \citet{evansetal09} quantify the observed inefficiency in terms of the star formation rate per free-fall time, defined as \citep{kt07}
\begin{equation}
SFR_{\rm ff} = \dot{M}_{*} t_{\rm ff} / M_{\rm cloud},
\end{equation}
where $t_{\rm ff}$ is defined as the free-fall time for the mean density of the cloud and which here we take as the initial free-fall time for our initially uniform density clouds.

Using this measure we find star formation rates of $SFR_{\rm ff} = 0.23$, $0.12$,  $0.15$ and $0.10$ for the four runs ($M/\Phi=\infty$, $10$, $5$ and $3$ respectively) that include radiative feedback and $SFR_{\rm ff} = 0.32$, $0.18$, $0.17$ and $0.12$ for the four runs using a barotropic equation of state, that is, neglecting radiative feedback. Thus, only the strong magnetic field calculations ($\beta < 1$, corresponding to $M/\Phi \lesssim 7$) that include radiative feedback approach the observed range of $SFR_{\rm ff} = 0.03-0.06$ measured by \citet{evansetal09}. All of the calculations with weaker field strengths and/or neglecting radiative feedback have star formation rates that are much higher than observations suggest. From the point of view of matching theory to observation, this is reassuring, since,} as discussed in \citet{pb08}, the most realistic of our calculations in terms of magnetic field strength is the strongest field case, $M/\Phi = 3$, since molecular cloud cores are typically observed with mass-to-flux ratios that are marginally supercritical (i.e. $M/\Phi \sim 2-3$) and with magnetic pressure smaller than gas pressure by a factor of $\sim 3$ (i.e., $\beta \sim 0.3$) \citep{crutcher99,ht04}. However we caution that any conclusions regarding the star formation efficiency from these calculations are necessarily limited by the relatively short period over which we have been able to follow the calculations beyond the free-fall time.

Furthermore our results present only a lower limit on the effect of feedback since we have neglected feedback from within 0.5~AU of a star including the driving of stellar winds and collimated outflows which may act to further reduce the star formation efficiency \citep{mm00}, perhaps explaining the remaining discrepancy between the efficiencies we find and the observed range of $3-6$\%.

\subsection{Implications for theory}

  The reduction in star formation rate is primarily a result of the support provided to the cloud by the magnetic field. The global magnetic field, whilst not sufficient to prevent collapse altogether, is nevertheless able to affect the binding energy. 
  
 \citet{cbk08} point out that the star formation rate can be made arbitrarily low in globally unbound clouds by increasing the ratio of kinetic to gravitational potential energy $E_{kin}/\vert E_{grav}\vert$ (set to unity in the initial conditions for the calculations we present here). The fact that increasing the turbulent velocity dispersion can decrease the efficiency of star formation in the sense of lowering the star formation rate has also been discussed previously \citep[e.g.][]{padoan95,khm00}. However, the kinetic energy cannot be increased indefinitely for a cloud of this size without violating the observational constraints on the turbulent velocity field.  Observationally, the velocity line width scales with cloud size approximately as $v \propto L^{0.5}$ \citep{larson81,solomonetal87,bh02,hb04} with a magnitude of $v\approx 1$~km~s$^{-1}$ on $1$~pc scales and a scatter of a factor of two \citep{hb04}.  For a cloud the size of those modelled here (0.375~pc), this gives a typical velocity dispersion of $v \approx 0.6$~km~s$^{-1}$ (Mach 3.3) which is almost a factor of two less than the velocity dispersion of our initial conditions. Thus, our initial conditions are already at the upper end of the observed velocity dispersion in molecular clouds so there would appear to be little scope for achieving a lower star formation rate by boosting the level of turbulence.

By contrast, as we have shown through the simulations presented here, a low star formation rate requires only a magnetic field of similar strength to observational estimates (i.e. a mass-to-flux ratio of $\gtrsim 3$, \citealt{crutcher99}) and the effects of radiative feedback which has no large free parameters (once the metallicity is set). Similar results with regards to the reduction in star formation rate with magnetic field strength are found by \citet{vsetal05} in the context of (scale-free) driven turbulence simulations.
 
 It therefore appears that both strong magnetic fields and radiative feedback from protostars are crucial ingredients in regulating star formation to a slow and inefficient level, which cannot be neglected from numerical simulations of the star formation process. 

\section*{Acknowledgments}
We thank the referee for a thorough, critical and helpful report. DJP would like to thank Christoph Federrath for useful discussions. DJP is currently supported by a Monash Fellowship, although part of this work has been completed whilst funded by a UK Royal Society University Research Fellowship. MRB is grateful for the support of a EURYI Award. This work, conducted as part of the award ``The formation of stars and planets: Radiation hydrodynamical and magnetohydrodynamical simulations"  made under the European Heads of Research Councils and European Science Foundation EURYI (European Young Investigator) Awards scheme, was supported by funds from the Participating Organisations of EURYI and the EC Sixth Framework Programme.  The calculations were performed on the University of Exeter Supercomputer, an SGI Altix ICE 8200. Figures containing images from the calculations were produced using SPLASH \citep{splashpaper}.

\appendix

\bibliography{sph,mhd,starformation}

\label{lastpage}
\enddocument